\documentclass[%
 aip,
 amsmath,amssymb,
 reprint,%
]{revtex4-1}

\usepackage{graphicx}
\usepackage{dcolumn}
\usepackage{bm}

\usepackage[utf8]{inputenc}
\usepackage[T1]{fontenc}
\usepackage{mathptmx}

\begin{document}

\preprint{AIP/123-QED}

\title[]{The stochastic model of neutrino oscillation \\}

\author{Xiao-Yan Wang}
 \altaffiliation[Also at ]{Department of Physics, Harbin Institute of Technology}
\author{Xiang-Jun Chen}%
 \email{chenxj@hit.edu.cn}
\affiliation{ Department of Physics, Harbin Institute of Technology
}%

\date{\today}

\begin{abstract}
Neutrino oscillation is phenomenon of random transition from a flavor state of neutrino to another, and should obey quantum statistics theory, and constitutes Markoffian process. The process is depicted by method of CTRW (continuous time random walk), and the time-evolution formula of flavor distribution of neutrino beam has been established. The simulation with Markoffian model for solar and cosmic ray neutrino indicates neutrino oscillation will arrive at equilibrium flavor distribution of three-fold maximum if propagation time is long enough, which is consistent to the data of SK and SNO experiments.
\end{abstract}

\maketitle

\section{\label{sec:level1} Introduction}

Neutrino oscillation is one of the greatest discoveries in particle physics for over twenty years, which was verified by SK Collaboration\cite{sk} in 1998 and confirmed by SNO\cite{sno} experiment in 2002. According to data of these two experiments, it was  indicated that the flavor probability distribution of beam neutrino would arrive at three-fold maximum, namely 1/3:1/3:1/3, when propagation time was long enough. It seemed that neutrino oscillation obeyed some statistical law. If we take the flavor state   as state variable of neutrino in beam (the value range only including three flavors $v_e, v_{\mu}, v_{\tau}$  ), neutrino beam will be a quantum ensemble, and its dynamics will obey quantum statistics theory. One  aim of the paper is to figure out whether the flavor kinetic equilibrium distribution of neutrino oscillation is three-fold maximum.

Quantum stochastic theory was first introduced into physics research by Einstein in 1905\cite{brown}, in order to study the motion of suspended particles in liquid. It was demonstrated in his paper that the random motion of suspended particles obeyed molecular-kinetic theory of heat, and constituted a diffusion process, and as a result particle probability density satisfied Gaussian distribution. We consider neutrino oscillation similar to the Brown motion of suspended particle from the property of motion: the Brown motion depicts the random continuous variation of particle position; neutrino oscillation depicts the random discrete variation of neutrino flavor state, which is result of the random transition from one flavor state of neutrino to another (one of three flavors$v_e, v_{\mu}, v_{\tau}$). Therefore we will investigate the statistical law of neutrino oscillation in a structure used by Einstein for suspended particle as follows: (1) on the aspect of dynamics, neutrino flavor conversion is something like propagator in flavor space in quantum mechanics, hence has Markoffian property, and constitutes Markoffian process. (2) on the aspect of probability distribution, according to quantum mechanics, neutrino flavor probability distribution obeys neutrino coherent theory. In Sec.2 we will elaborate the quantum stochastic process of neutrino oscillation directly by investigating the evolution of flavor distribution of neutrino beam, which is the statistical law of neutrinos in beam.

 Quantum stochastic theory is applied widely in fields of applied physics like optical material\cite{ctrw,current,effect} and so on. While to the best of our knowledge, it appears that quantum stochastic theory has not been applied in particle physics in the literature. The purpose of this work is to address this issue and derive the stochastic process of neutrino oscillation and establish corresponding stochastic model. The Markoffian process of neutrino oscillation will be depicted by a method called CTRW\cite{ctrw,weak,from} (continuous time random walk), and neutrino flavor probability will be calculated according to neutrino coherent theory, which can avoid the complex calculation of concrete interaction Hamiltonian. For convenience, we will discretize time and establish the homogeneous time-discrete Markoffian model. The time-evolution of flavor probability distribution in neutrino oscillation will be simulated by the Markoffian model to determine the equilibrium flavor distribution of neutrino oscillation. 

Due to the long lifetime of neutrino\cite{decay}, we assume the effect of medium is the only factor which will influence neutrino oscillation. For neutrino is neutral particle with tiny mass and strong penetration, the influence of medium effect on neutrino oscillation is feeble and we will discuss it in the concrete cases. 

This paper will be constructed as follows: In Sec.2, we will demonstrate the underlying Markoffian process of neutrino oscillation, derive the time-evolution equation of flavor distribution during neutrino beam propagation, and discretize time and establish corresponding homogeneous time-discrete Markoffian model; Sec.3,  we will evolve our Markoffian model to simulate the evolution of flavor distributions of solar and cosmic ray neutrino, and estimate the distributions on the earth's surface of these two cases; Sec.4, we will make conclusions.

\section{\label{sec:level2}Markoffian model}
\subsection{\label{sec:level3}Markoffian process of neutrino oscillation}

The Hamiltonian of neutrino oscillation could be written as 

\begin{eqnarray}
H&=&H_0 + H_I  \nonumber \\
&=&\hat{ p}^2/2m + \frac{G_F'}{\sqrt{2}}\bigg(\sum_{i,j} \bar v_{jL} \gamma^{\mu}v_{iL}\bigg)\bigg(\sum_{k,l} \bar v_{kL}\gamma_{\mu}v_{lL}\bigg) 
\end{eqnarray}
Here, $H_0$ is the kinetic energy which represents neutrino beam's translation;  $H_I$ represents interactions by which neutrino flavor conversions are induced. Dependent on the choice of  $H_I$, the process of neutrino oscillation may be stochastic process or not. From the expression of interaction (the second term on the right of the second equation), it is explicit that the initial flavor state $v_i (or v_l) $ of neutrino randomly transits to any of three flavor states  $v_e, v_{\mu}, v_{\tau}$ . Thus, $H_I$ expresses random flavor conversion and neutrino oscillation is stochastic process.

Consider neutrino beam including N particles which labeled by numbers 1,2,$\cdots$,N. The quantum state of particle K is denoted as $|K> (K=1,2,\cdots, N) $. Then, the density operator of neutrino beam  is

\begin{eqnarray}
\hat{\rho} \equiv \frac{1}{N} \sum_K |K><K| 
\end{eqnarray}

 In this paper, we adopt three-flavor neutrino frame. Thus three neutrino flavor states $|v_e>, |v_{\mu}>, |v_{\tau}>  $ span neutrino flavor Hilbert space, in other words, these three flavor vectors constitute a complete set of orthonormal basis in flavor space. The state  $|K>$ is expanded by these three flavor basis vectors as follows:

\begin{eqnarray}
|K> = \sum_i <v_i | K > |v_i> =\sum_i C_{iK} |v_i>    \quad (i=e,\mu,\tau)
\end{eqnarray}
The coefficient $ C_{iK}=<v_i|K>  $ expresses the $v_i$ component wave function in flavor representation, which represents the probability amplitude of particle K staying at flavor state $|v_i>$.

The density matrix element in flavor representation is

\begin{eqnarray}
\rho_{ij} = \frac{1}{N} \sum_K <v_i|K><K| v_j>     \quad (i,j=e,\mu ,\tau )
\end{eqnarray}
and the diagonal element

\begin{eqnarray}
\rho_{ii} = \frac{1}{N} \sum_K <v_i|K><K| v_i> =  \frac{1}{N} \sum_K |<v_i|K>|^2 = \frac{1}{N}\sum_K |C_{iK}|^2 \nonumber \\
\end{eqnarray}
expresses the average probability of neutrino in beam staying at state $|v_i>$, in other words, $\rho_{ii}$  is the flavor probability of $|v_i>$ in neutrino beam. Thus the flavor probability distribution can be written as $\rho_{ee}:\rho_{\mu \mu} :\rho_{\tau \tau}    $.

The purpose of this paper is to investigate the time-evolution of flavor distribution of neutrino beam. Hence we only need to study the time-evolution of diagonal elements.

Given the density operator of beam $\hat{\rho}(t_0)$ at time $t_0 $, in Heisenberg picture, the time-evolution of the density operator is

\begin{eqnarray}
\hat{\rho}(t) = e^{i\hat{H}(t-t_0)} \hat{\rho}(t_0) e^{-i\hat{H}(t-t_0)  }
\end{eqnarray}
In flavor representation, the time-evolution of diagonal element is

\begin{eqnarray}
<n|\hat{\rho}(t)|n> &=&<n| e^{i\hat{H}(t-t_0)} \hat{\rho}(t_0) e^{-i\hat{H}(t-t_0)  }|n> \nonumber \\
&=&\sum_i <n| e^{i\hat{H}(t-t_0)}|i><i| \hat{\rho}(t_0) \sum_j |j><j|e^{-i\hat{H}(t-t_0)  }|n> \nonumber\\
\rightarrow \rho_{nn}(t) &=&\sum_{ij}  <n| e^{i\hat{H}(t-t_0)}|i><j|e^{-i\hat{H}(t-t_0)  }|n> \rho_{ij}(t_0) \nonumber  \\
&=& \sum_i |<n| e^{i\hat{H}(t-t_0)}|i>|^2 \rho_{ii}(t_0)  \nonumber \\
 &&+ \sum_{i\ne j} <n| e^{i\hat{H}(t-t_0)}|i><j|e^{-i\hat{H}(t-t_0)  }|n>\rho_{ij}(t_0)
\end{eqnarray}
Here, we have employed the identical equation

\begin{eqnarray}
\sum_i |v_i><v_i| = I   \quad (i=e,\mu,\tau)   \nonumber
\end{eqnarray}
In formula (7), the left of the last equation is flavor probability of $|v_n>$ at time t; the first term on the right expresses total transition probability of all flavor states at time $t_0$ transiting to  $|v_n>$ at t, and the second term on the right can be considered as perturbation term, which can be omitted at zero order approximation compared with the first term. Thus the time-evolution formula of flavor distribution of neutrino beam is:

\begin{eqnarray}
\rho_{nn}(t)\approx \sum_i \big|<n| e^{i\hat{H}(t-t_0)}|i>\big|^2 \rho_{ii}(t_0) = \sum_i\big|K_{n,i}(t,t_0)\big|^2 \rho_{ii}(t_0) \nonumber \\
\end{eqnarray}
where $K_{n,i}(t,t_0)= <n| e^{i\hat{H}(t-t_0)}|i>   $  is propagation(conversion) probability amplitude of from state $|v_i>$ to state  $|v_n>$. Formula (8) expresses that the flavor probability of  $|v_n>$ at time t equals total transition probability of all flavor states at time   $t_0$ transiting to  $|v_n>$ at t. Thus formula (8) is the form of Markoffian conditions (formula 2.3.3 in Sec.2 in literature\cite{statistical}) in flavor representation.

 Formula (8) can be rewritten as

\begin{eqnarray}
\rho_{nn}(t)&= & \sum_i\big|K_{n,i}(t,t_1)\big|^2 \sum_j\big|K_{i,j}(t_1,t_0)\big|^2 \rho_{jj}(t_0) \nonumber \\
&=&\sum_j\big|K_{n,j}(t,t_0)\big|^2 \rho_{jj}(t_0)
\end{eqnarray}
Here, $ \rho_{jj}(t_0) $  represents flavor probability of $|v_j>$ at time $t_0$. Thus

\begin{eqnarray}
\big|K_{n,j}(t,t_0)\big|^2 =  \sum_i\big|K_{n,i}(t,t_1)\big|^2 \big|K_{i,j}(t_1,t_0)\big|^2
\end{eqnarray}
is established. Obviously, formula (10) is the form of Markoffian conditions (formula 1.1.22 in Sec.2 in literature\cite{statistical}) in flavor representation.

For propagation probability amplitude $K_{n,j}(t,t_0)$,

\begin{eqnarray}
 \sum_j\big|K_{n,j}(t,t_0)\big|^2 )=  \sum_j\big| <n| e^{i\hat{H}(t-t_0)}|j> \big|^2=1
\end{eqnarray}
is established for particle number conservation. (Formula (11) expresses the total transition probability of all neutrino flavor states at  $t_0$ all transiting to state  $|v_n>$ at t.)

Now, we will use propagation probability distribution $ \big|K(t,t_0)\big|^2 $ to construct Markoffian process of neutrino oscillation in flavor space, and will describe it by method of CTRW (continuous time random walk) as follows: consider one walker at location $x_0$ in flavor space at time $t_0$, and time is discretized by interval $\Delta t$.(Location $x_0$  is determined by three flavor probabilities $\big(|\psi_{v_e}(t_0)|^2, |\psi_{v_{\mu}}(t_0)|^2, |\psi_{v_{\tau}}(t_0)|^2        \big)   $ at  $t_0$, and according to formula (5) $ |\psi_{v_i}(t_0)|^2 = \rho_{ii}=\frac{1}{N}\sum_K |C_{iK}|^2, i=e,\mu, \tau  $.) The walker starts from position  $x_0$  at time $t_0$, after interval $\Delta t$, moves to a new position $x_1$ with propagation probability distribution $ \big|K_{1,0}(\Delta t)\big|^2 $; then starts from  $x_1$, and moves to  $x_2$ after interval  $\Delta t$ with propagation probability distribution $ \big|K_{2,1}(\Delta t)\big|^2 $, $\cdots$.Thus the Markoffian process of neutrino oscillation is constructed, with propagation probability distribution  $ \big|K_{n,m}(\Delta t)\big|^2 $

\begin{eqnarray}
 \big|K_{n,m}(\Delta t)\big|^2 )= \big| _{t_n}<v_j| e^{i\hat{H}(\Delta t)}|v_i>_{t_m} \big|^2  \nonumber  \\
(i,j =e,\mu, \tau ;\ \Delta t = t_n-t_m   )
\end{eqnarray}
Here,$ |v_i>_{t_m} $ is any flavor state at time $t_m$ and $ |v_j>_{t_n} $ is any flavor state at time $t_n$; so formula (12) expresses all possible conversion probabilities of from flavor state at time  $t_m$ jumping once to flavor state at time $t_n$ , namely propagation probability distribution.

\subsection{\label{sec:level4} The time-evolution of flavor probability distribution of neutrino beam }

Now we will investigate the evolution of flavor distribution after the walker jump once. To avoid calculating complex interaction Hamiltonian $H_I  $, the flavor probability will be calculated according to neutrino coherent theory, ignoring details of interaction.

In neutrino coherent theory, formulas of neutrino flavor conversion probability are obtained by two-flavor mixing approximation. Given the initial neutrino flavor $v_{\alpha}$ and possible final states of conversion $v_{\beta},v_{\gamma}\quad (\alpha \ne \beta \ne \gamma)$, formulas of neutrino flavor conversion probability will be respectively

\begin{eqnarray}
P_0(v_{\alpha}\rightarrow v_{\beta})&= &sin^2(2\theta_{ij}) sin^2 \big( \frac{\Delta  m_{ij}^2}{4E} t \big)\quad(i \neq j)     \\
P_0(v_{\alpha}\rightarrow v_{\gamma})&=& sin^2(2\theta_{ik}) sin^2 \big( \frac{\Delta  m_{ik}^2}{4E} t \big)\quad (i \neq k)
\end{eqnarray}
and the survival probabilities of initial flavor will respectively be

\begin{eqnarray}
P_0(v_{\alpha}\rightarrow v_{\alpha} )&=&1-P(v_{\alpha}\rightarrow v_{\beta})=1- sin^2(2\theta_{ij}) sin^2 \big( \frac{\Delta  m_{ij}^2}{4E} t  \big) \nonumber \\    \\
P'_0(v_{\alpha}\rightarrow v_{\alpha})&=&1-P(v_{\alpha}\rightarrow v_{\gamma})=1- sin^2(2\theta_{ik}) sin^2 \big(\frac{\Delta  m_{ik}^2}{4E} t  \big)  \nonumber \\
\end{eqnarray}

In three-flavor frame, we assume the statuses of two conversion modes $v_{\alpha}\rightarrow v_{\beta}$ and $v_{\alpha}\rightarrow v_{\gamma}$ are equal, then the survival probability of initial flavor  $v_{\alpha}$ should be

\begin{eqnarray}
P(v_{\alpha}\rightarrow v_{\alpha} )&=&\frac{P_0(v_{\alpha}\rightarrow v_{\alpha} )+P'_0(v_{\alpha}\rightarrow v_{\alpha})}{2} \nonumber \\
&=&1-\frac{P_0(v_{\alpha}\rightarrow v_{\beta})+P_0(v_{\alpha}\rightarrow v_{\gamma})}{2}  \nonumber \\
&=&1- \frac{1}{2}\bigg[sin^2(2\theta_{ij}) sin^2 \big( \frac{\Delta  m_{ij}^2}{4E} t  \big)        + sin^2(2\theta_{ik}) sin^2 \big( \frac{\Delta  m_{ik}^2}{4E} t \big)\bigg] \nonumber \\     
\end{eqnarray}
and flavor conversion probabilities should be correspondingly

\begin{eqnarray}
P(v_{\alpha}\rightarrow v_{\beta})= \frac{1}{2}sin^2(2\theta_{ij}) sin^2 \big( \frac{\Delta  m_{ij}^2}{4E} t \big) \\
P(v_{\alpha}\rightarrow v_{\gamma})=\frac{1}{2} sin^2(2\theta_{ik}) sin^2 \big( \frac{\Delta  m_{ik}^2}{4E} t \big)
\end{eqnarray}

 Given the flavor distribution at time $t_0$, at position  $x_0$ is

\begin{eqnarray}
P_e(x_0,t_0):P_{\mu}(x_0,t_0):P_{\tau}(x_0,t_0)
\end{eqnarray}
A walker jumps once from position $ (x_0,t_0)  $ to $ (x,t)$, and the probability of any neutrino flavor like $v_e$ at position $ (x,t)$, $P_e(x,t)  $, is composed of three parts: (1), the survival probability of flavor $v_e$ with probability $ P_e(x_0,t_0) $ at  $ (x_0,t_0)  $ , and the calculation formula (17); (2), the probability obtained from flavor conversion of the other two flavors $v_{\mu}, v_{\tau}$ with their probability $ P_{\mu}(x_0,t_0), P_{\tau}(x_0,t_0) $ at position $ (x_0,t_0)  $, and the calculation formulas (18) and (19):

\begin{eqnarray}
P_e(x,t)&=&P_e(x_0,t_0)P_{e\rightarrow e}(\Delta t)+ P_{\mu}(x_0,t_0)P_{\mu \rightarrow e}(\Delta t)+ P_{\tau}(x_0,t_0)P_{\tau \rightarrow e}(\Delta t)  \nonumber \\
&=&\left(\begin{array}{ccc} P_e(x_0,t_0) &  P_{\mu}(x_0,t_0) & P_{\tau}(x_0,t_0) \end{array} \right)  \left(\begin{array}{c} P_{11}(\Delta t) \\ P_{21}(\Delta t) \\  P_{31}(\Delta t)  \end{array} \right)
\end{eqnarray}

The calculation of the other two flavor probabilities at position $ (x,t)$ is similar. Thus the evolution formula of flavor distribution after walker jump once is:

\begin{eqnarray}
&&\left(\begin{array}{ccc} P_e(x,t) &  P_{\mu}(x,t) &  P_{\tau}(x,t) \end{array} \right)   \nonumber \\
&=&\left(\begin{array}{ccc} P_e(x_0,t_0) &  P_{\mu}(x_0,t_0) & P_{\tau}(x_0,t_0) \end{array} \right)  \left(\begin{array}{ccc} P_{11}(\Delta t) &  P_{12}(\Delta t) &  P_{13}(\Delta t)  \\
 P_{21}(\Delta t) &  P_{22}(\Delta t) &  P_{23}(\Delta t)\\ 
 P_{31}(\Delta t) &  P_{32}(\Delta t) & P_{33}(\Delta t)  \end{array} \right) \nonumber  \\ 
\end{eqnarray}
Obviously, formula (22) is the evolution formula of flavor distribution after Markoffian process evolving over $\Delta t  $.

\subsection{\label{sec:level5} Time-discrete Markoffian model}

For Markoffian model employed conveniently, we should discretize time properly and construct time-discrete Markoffian model. The problem is how to choose appropriate time interval (denoted as $\Delta t_T $ ) to make Markoffian model simplest, in other words make the transition matrix (22) independent of time, namely homogeneous Markoffian model.

For the time-dependent terms in transition matrix(22) are all trigonometric functions, the time could be eliminated by making all trigonometric functions extreme simultaneously. Time-dependent trigonometric functions in transition matrix (22) are listed below:

\begin{eqnarray}
sin^2 \big( \frac{\Delta  m_{12}^2}{4E}\Delta t \big),\quad sin^2 \big( \frac{\Delta  m_{13}^2}{4E}\Delta t \big),\quad sin^2 \big( \frac{\Delta  m_{23}^2}{4E}\Delta t \big)
\end{eqnarray}
Thus, the problem to choose proper interval $\Delta t_T $ to make transition matrix(22) timeless, is equivalent to choose interval  $\Delta t_T $ to make three trigonometric functions in (23) extreme simultaneously. For more precise simulation with our model, it is required that the interval $\Delta t_T $ should be as small as possible. Given the first two terms in (23) 
\begin{eqnarray}
sin^2 \big( \frac{\Delta  m_{12}^2}{4E} \Delta t \big),\quad sin^2 \big( \frac{\Delta  m_{13}^2}{4E}\Delta t \big)  \nonumber
\end{eqnarray}
both arrive at maximum by the smallest interval $\Delta t $, if the third term
\begin{eqnarray}
 sin^2 \big( \frac{\Delta  m_{23}^2}{4E}\Delta t \big) \nonumber
\end{eqnarray}
arrives at extreme too, the interval $\Delta t $ would be $\Delta t_T $ we want. Considering the mass of $v_e$ close to  $v_{\mu}$, $|\Delta  m_{13}^2| \sim |\Delta  m_{23}^2 | $. Thus

\begin{eqnarray}
\frac{|\Delta  m_{13}^2|}{4E}   \sim \frac{|\Delta  m_{23}^2|}{4E}
\end{eqnarray}
and the interval  $\Delta t $  maximizes $sin^2 \big( \frac{\Delta  m_{13}^2}{4E}\Delta t \big)$ will approximately maximize $  sin^2 \big( \frac{\Delta  m_{23}^2}{4E}\Delta t \big) $.Thus, when Markoffian model discretized by the interval $\Delta t $, transition matrix (22) will become time-independent as follows

\begin{equation}
\frac{\nu_{e}\qquad \qquad  \nu_{\mu}\qquad \qquad \nu_{\tau} }
{\mathbf{P}=\left(\begin{array}{c} \\ \nu_{e} \\ \\ \nu_{\mu} \\ \\
\nu_{\tau} \end{array} \right)
 \left( \begin{array}{ccc} & & \\
 p_{11} & \frac{1}{2}{\tiny \sin^2(2\theta_{12})} & \frac{1}{2}{\tiny \sin^2(2\theta_{13})} \\
 & & \\
\frac{1}{2}{\tiny \sin^2(2\theta_{21})} & p_{22} & \frac{1}{2} {\tiny \sin^2(2\theta_{23})} \\
 & & \\
 \frac{1}{2}{\tiny \sin^2(2\theta_{31})} & \frac{1}{2} {\tiny
\sin^2(2\theta_{32})} & p_{33}
 \end{array} \right)}
\end{equation}
where $ p_{ii} = 1-\Sigma_{i\ne j} p_{ij}  $ is the survival probability of $v_i~ (i=e,\mu, \tau)$.And Markoffian model will become homogeneous time-discrete. 

It is worth to mention that formula (17) will arrive at the first lowest value by the interval  $\Delta t_T $. Thus the step length of Markoffian model is the distance over which the survival probability of initial neutrino flavor arrives at the first lowest value. According to experimental data, the step length  is at most 295km\cite{t2k}, which obtained when the initial neutrino flavor is $v_{\mu}$.

\section{\label{sec:level6}Simulation with Markoffian model}
\subsection{\label{sec:level7} Evolution of Markoffian model   }

Now we will evolve the time-discrete Markoffian model to simulate the time-evolution of flavor distribution during neutrino beam propagation and investigate the flavor distributions of solar and cosmic ray neutrino on the earth's surface. 

Taking the global best-fit values of neutrino mixing angles $ \theta_{12}\simeq 34^o, \theta_{23}\simeq 45^o, \theta_{13}\simeq 10^o $   into (25), and the numerical one-step transition matrix in vacuum is

\begin{equation}
\mathbf{P}= \left( \begin{array}{ccc}
 0.51 & 0.43 & 0.06 \\
 0.43 & 0.07 & 0.50 \\
 0.06 & 0.50 & 0.44
 \end{array} \right)
\end{equation}

The evolution of Markoffian model is to multiply the flavor probability distribution with one-step transition matrix (formula 22). For example, given the initial flavor distribution ($I_e : I_{\mu}:I_{\tau}  $ ), the evolution of flavor distribution for Markoffian model evolving one step is 

\begin{equation}
 ( I_e \quad I_{\mu} \quad I_{\tau} ) P=  ( I_e \quad I_{\mu} \quad I_{\tau} )\left( \begin{array}{ccc}
 0.51 & 0.43 & 0.06 \\
 0.43 & 0.07 & 0.50 \\
 0.06 & 0.50 & 0.44
 \end{array} \right)
\end{equation}
and two steps 

\begin{equation}
 ( I_e \quad I_{\mu} \quad I_{\tau} ) P^2=  ( I_e \quad I_{\mu} \quad I_{\tau} )\left( \begin{array}{ccc}
 0.51 & 0.43 & 0.06 \\
 0.43 & 0.07 & 0.50 \\
 0.06 & 0.50 & 0.44
 \end{array} \right)\left( \begin{array}{ccc}
 0.51 & 0.43 & 0.06 \\
 0.43 & 0.07 & 0.50 \\
 0.06 & 0.50 & 0.44
 \end{array} \right)
\end{equation}

\subsection{\label{sec:level8} The equilibrium flavor distribution of solar neutrino oscillation and the flavor distribution on the earth's surface }

Solar neutrino's propagation from the Sun to the earth’s surface should experiences two different stages: The first stage insider the Sun is from the solar core to the solar surface, where neutrino oscillation is influenced by MSW effect; the second stage is from the solar surface to the earth's surface, where oscillation approximately proceeds in vacuum. 

Due to the solar interior of continuously variable density, when neutrino propagates insider the sun,  neutrino oscillation with different energy will be influenced differently by MSW effect. According to literature\cite{msw}, for neutrino energy $ > $ 10MeV (high energy neutrino), neutrino oscillation is suppressed and flavor conversion is an adiabatic process. For this case, we adopt the flavor distribution on the solar surface as the initial distribution for the evolution of Markoffian model. And according to literature\cite{msw}, when high energy neutrino propagates to the solar surface undergoing adiabatic flavor conversion, the survival probability of $v_e$   on the solar surface would be $ sin^2\theta_{12} $, $ v_{\mu}  $   and $ v_{\tau}  $ share the rest probability. According to muon neutrino oscillation experiments\cite{sk,t2k}, the mixing between   $ v_{\mu}  $   and $ v_{\tau}  $   is maximal and we assume  $ v_{\mu}  $,   $ v_{\tau}  $  of high energy equally share the rest probability. Thus the flavor distribution of high energy neutrino on the solar surface, denoted as $I_1$, is

\begin{equation}
I_1 =  v_e : v_{\mu} : v_{\tau} \approx sin^2 \theta_{12} : \frac{1}{2} cos^2 \theta_{12} : \frac{1}{2} cos^2 \theta_{12}
\end{equation}
Take the best-fit value of $\theta_{12} =34^o  $  into (29) and obtain

\begin{equation}
I_1= v_e : v_{\mu} : v_{\tau} = 0.3126 : 0.3437 : 0.3437
\end{equation}

For neutrino of energy $<$ 10MeV (low energy neutrino), oscillation will not be suppressed by MSW effect\cite{msw} insider the sun. For the small medium effect, we will ignore it in the case. The distance from the solar core to the earth's surface is the total distance over which Markoffian model evolves. And the initial distribution is the distribution at the solar core, denoted as $I_2$

\begin{equation}
I_2= v_e : v_{\mu} : v_{\tau} = 1 : 0: 0
\end{equation}

The aim of this subsection is to investigate the flavor distribution of solar neutrino on the earth's surface, which should be the total flavor distribution of above two kinds of solar neutrino. Thus the total distribution on the solar surface is the initial distribution from which Markoffian model evolves starting.

Suppose the flavor distribution of low energy neutrino on the solar surface is $I_2'=I_2 P^{n'}$, and the total flavor distribution on the solar surface is the linear combination of $I_1, I_2'$ .

\begin{equation}
I = a I_1 + b I'_2, \qquad  a+b =1
\end{equation}

The  resultant distribution after the Markoffian model evolves with constant transition matrix (26) is

\begin{equation}
(aI_1 +b I'_2)P^n =a I_1 P^n + b I'_2 P^n = a I_1^{(n)} + b I_2^{(n+n')}
\end{equation}

For neutrino of energy $ >$ 10MeV , the flavor distributions for the evolution of the Markoffian model are shown in Table-1. And the distributions for neutrino of  energy $<$ 10MeV are shown in Table-2.

\begin{table}
\caption{\label{tab:table1}Simulation of solar neutrino of energy $>$ 10MeV }
\begin{ruledtabular}
\begin{tabular}{cc}
Steps of  evolution   & Flavor distribution ( $ v_e : v_{\mu} : v_{\tau} $) \\
\hline
0 & 0.3126 \quad 0.3437 \quad 0.3437 \\
\hline
1 &  0.3278  \quad  0.3303  \quad  0.3418 \\
\hline
2 &  0.3297  \quad   0.3350 \quad   0.3352 \\
\hline
3 & 0.3323 \quad   0.3329 \quad   0.3348 \\
\hline
4 &  0.3327 \quad   0.3336  \quad    0.3337 \\
\hline
5 & 0.3332  \quad  0.3333 \quad   0.3336 \\
\end{tabular}
\end{ruledtabular}
\end{table}

\begin{table}
\caption{\label{tab:table2}Simulation of solar neutrino of energy $<$ 10MeV }
\begin{ruledtabular}
\begin{tabular}{cc}
Steps of evolution  & Flavor distribution ( $ v_e : v_{\mu} : v_{\tau} $) \\
\hline
0 &  1 \quad 0 \quad 0 \\
\hline
1 & 0.5100 \quad   0.4300 \quad   0.0600 \\
\hline
2 & 0.4486 \quad   0.2794  \quad  0.2720 \\
\hline
3 & 0.3652  \quad  0.3485  \quad  0.2863 \\
\hline
4 &  0.3533  \quad  0.3246 \quad   0.3221 \\
\hline
5 & 0.3391 \quad   0.3357   \quad  0.3252 \\
\hline
6 & 0.3368  \quad  0.3319  \quad  0.3313 \\
\end{tabular}
\end{ruledtabular}
\end{table}

The  equilibrium flavor distributions shown in Table-1 and Table-2 are both
\begin{eqnarray}
I^{(e)} = 1/3 : 1/3 : 1/3
\end{eqnarray}
So, when the flavor distributions of these two cases both arrive at the equilibrium distribution, the values of coefficients a, b in (33) become insignificant. 

The distributions in Table-1 and Table-2 indicate that after the Markoffian model evolves for over 6 steps, the distributions of two kinds of solar neutrino will both arrive at the equilibrium distribution of three-fold maximum. As mentioned in Sec.2, the step length of Markoffian model is 295km at the most, thus the distance between the sun and the earth is long enough for the Markoffian model evolving for over 6 steps. This means that the distribution of solar neutrino on the earth’s surface is the equilibrium distribution of three-fold maximum, which is just the result of SNO in 2002\cite{sno}.

\subsection{\label{sec:level9}The equilibrium flavor distribution of cosmic ray neutrino oscillation and the distribution after neutrino passing through the earth }

For neutrino in cosmic ray, the initial flavor distribution before neutrino passing through the earth is

\begin{equation}
 v_e : v_{\mu} : v_{\tau} = \frac{1}{3} : \frac{2}{3} : 0
\end{equation}

The distributions for the evolution  of the Markoffian model are shown in Table-3.

\begin{table}
\caption{\label{tab:table3}Simulation of neutrino in cosmic ray }
\begin{ruledtabular}
\begin{tabular}{cc}
Steps of evolution & Flavor distribution  ( $ v_e : v_{\mu} : v_{\tau} $) \\
\hline
0 & 1/3 : 2/3 : 0 \\
\hline
1 & 0.4567 : 0.1900 : 0.3533 \\
\hline
2 & 0.3358 : 0.3863 : 0.2779 \\
\hline
3 & 0.3541 : 0.3104 : 0.3356 \\
\end{tabular}
\end{ruledtabular}
\end{table}

The equilibrium flavor distribution shown in Table-3 is three-fold maximum occurring after the Markoffian model evolving for over 3 steps, which is the result of SK in 1998\cite{sk}. (The diameter of the Earth is long enough for the Markoffian model evolving for over 3 steps.)

In summary, the three-fold maximum flavor distribution which indicated in SNO and SK experiments, is the equilibrium distribution of Markoffian process of neutrino oscillation.

\section{\label{sec:level10} Conclusions }

 Neutrino oscillation is random transition from one flavor state of neutrino to another and obeys quantum statistics theory. According to the theory,  the evolution of flavor distribution of neutrino beam satisfies Markoffian conditions. Markoffian process of neutrino oscillation is depicted by method of CTRW(continuous time random walk), and neutrino flavor probability is calculated with formulas in neutrino coherent theory, and then the time-evolution formula of flavor probability distribution (formula 22) is established.

For Markofffian model employed conveniently, the process is time-discretized and homogeneous time-discrete Markoffian model is established. The simulation with Markoffian model indicates that neutrino oscillation will arrive at equilibrium flavor distribution of three-fold maximum when propagation time is long enough. The flavor distributions of solar and cosmic ray neutrino on the earth's surface are both equilibrium flavor distribution of three-fold maximum, which is consistent to experimental data. 

\begin{acknowledgments}
We would like to thank Professor  Guo-Ping Du for useful discussions at Nanchang University, and thank Professor Gui-Xin Tang at Harbin Institute of Technology.
\end{acknowledgments}

\nocite{*}
\bibliography{markovbib}

\end{document}